\documentclass{article}
\usepackage{ijcai24}

\usepackage{times}
\usepackage{soul}
\usepackage{url}
\usepackage[hidelinks]{hyperref}
\usepackage[utf8]{inputenc}
\usepackage[small]{caption}
\usepackage{graphicx}
\usepackage{amsmath}
\usepackage{amsthm}
\usepackage{booktabs}
\usepackage{algorithm}
\usepackage{algorithmic}
\usepackage{enumitem}  
\usepackage[table]{xcolor}
\usepackage[switch]{lineno}
\usepackage{multirow}
\usepackage[T1]{fontenc}

\title{Impact of Noisy Labels on Sound Event Detection: Deletion Errors Are More Detrimental Than Insertion Errors}

\author{
Yuliang Zhang
\and
Roberto Togneri \and
Defeng (David) Huang\\
\affiliations
School of Electrical, Electronic and Computer Engineering, The University of Western Australia\\
\emails
\ yuliang.zhang@research.uwa.edu.au,
\{roberto.togneri, david.huang\}@uwa.edu.au
}

\begin{document}

\maketitle

\begin{abstract}

    This study explores the critical but underexamined impact of label noise on Sound Event Detection (SED), which requires both sound identification and precise temporal localization. We categorize label noise into deletion, insertion, substitution, and subjective types and systematically evaluate their effects on SED using synthetic and real-life datasets. Our analysis shows that deletion noise significantly degrades performance, while insertion noise is relatively benign. Moreover, loss functions effective against classification noise do not perform well for SED due to intra-class imbalance between foreground sound events and background sounds. We demonstrate that loss functions designed to address data imbalance in SED can effectively reduce the impact of noisy labels on system performance. For instance, halving the weight of background sounds in a synthetic dataset improved macro-F1 and micro-F1 scores by approximately $9\%$ with minimal Error Rate increase, with consistent results in real-life datasets. This research highlights the nuanced effects of noisy labels on SED systems and provides practical strategies to enhance model robustness, which are pivotal for both constructing new SED datasets and improving model performance, including efficient utilization of soft and crowdsourced labels.

\end{abstract}

\section{Introduction}

The advent of deep learning has revolutionized numerous domains, such as computer vision \cite{c1} \cite{c2}, natural language processing \cite{c3} \cite{c4} \cite{c5}, and audio signal processing \cite{c32} \cite{c7} \cite{c8}, driven by the availability of large datasets with high-quality labels. Acquiring such datasets, however, is often an expensive and time-consuming endeavor. To reduce annotation costs, researchers have increasingly turned to the use of data annotated by ordinary people instead of experts, such as using crowdsourcing \cite{c9} \cite{c10} to acquire large amounts of data. The use of these sources, however, often results in unreliable labels, called noisy labels. Moreover, the data labels can be subjective even for experienced domain experts \cite{c11} \cite{c12}. 

Training  Deep Neural Networks (DNNs) with data that contains noisy labels usually leads to a significant decline in the models' performance \cite{c25}. The numerous parameters make the DNNs easy to overfit on error labels \cite{c13} \cite{c14}. To mitigate the effects of noisy labels, a variety of deep-learning approaches have been developed in recent years. The paper by \cite{c16} has classified these approaches into five distinct categories based on their methodological differences. However, the majority of these methods primarily address classification tasks, particularly in the realm of computer vision. In the field of Sound Event Detection (SED), the challenges are compounded as the task extends beyond mere identification of sound events to also include the precise determination of their onset and offset boundaries. This additional requirement exacerbates the difficulties posed by noisy labels. 

The annotation of SED has a very specific requirement referred to as \textit{strong labels}, where the annotation contains temporal information for each sound event instance, namely, its onset and offset times \cite{c26}. Due to the intricate nature of acoustic signals, such as polyphonic environments and low Signal-to-Noise Ratio (SNR), and the subjective judgment of annotators, labels for SED are more prone to noise introduction. Based on the experiments in \cite{c33}, a maximum of 80\% coincidence with the complete and correct reference annotations was obtained for synthetic data. However, the coincidence rate with reference labels was below 69\% for real data, indicating that noisy labels in sound event detection are a serious issue. 

Some researchers, like those in \cite{c10} and \cite{c30}, have focused on the problem of noisy labels in the sound event field. However, their efforts are still centered on mitigating the impact of label noise in classification tasks. In SED, there is a lack of systematic analysis concerning the different types of noisy labels and their impacts on system performance.

In this paper, we begin by dividing the label noise in SED data into four categories:
\begin{itemize}[parsep=0pt]
   \item \textbf{Deletion Label Noise:}
    Sound events that are overlooked or have underestimated durations. 
  \item \textbf{Insertion Label Noise:}
    Non-occurring sound events are mistakenly marked, or sound event durations are overestimated.
  \item \textbf{Substitution Label Noise:}
    A sound event is incorrectly labeled as another type of sound event.
  \item \textbf{Subjective Label Noise:}
    The variabilities among different annotators for the same audio segment. 
\end{itemize}  
Our research then focuses on examining how these various types of label noise affect the performance of SED models. Are some types of noise more disruptive than others? To address this, we explore three well-known noise-robust loss functions originally designed for multi-class classification tasks—soft bootstrapping \cite{c28}, label smoothing \cite{c24}, and generalized cross-entropy loss \cite{c29}—and extend them to multi-label classification to assess their effectiveness in mitigating these inaccuracies in SED. Finally, we present our key findings, outlining practical strategies that can be employed to counteract the challenges posed by label noise in SED systems. The contributions of this paper are summarized as follows:

\begin{enumerate}[label=\arabic*)] 
    \item To our knowledge, this study is the first to systematically examine the impact of various types of noisy labels on Sound Event Detection (SED) tasks. We provide practical methods for constructing datasets with noisy labels in both synthetic and real-life settings to assess their impact. We discovered that deletion label noise adversely impacts the system more than insertion label noise, as demonstrated by both experimental results and theoretical analysis.

    \item Loss functions effective against classification noise do not work well for SED due to the unique characteristics of intra-class imbalance between foreground sound events and background sounds. We demonstrated that loss functions designed to counter data imbalance in SED can effectively reduce the impact of noisy labels on system performance. A simple reweighting function was used to control the weight for foreground and background sounds separately. Experimental results showed that halving the weight of background sounds in a synthetic dataset raised macro-F1 and micro-F1 scores by approximately $9\%$, with minimal Error Rate increase. These improvements were consistent in the real-life dataset.

    \item This research highlights the nuanced effects of noisy labels on SED systems and provides practical strategies to enhance model robustness. These strategies are crucial for both constructing new SED datasets and improving model performance, particularly in the efficient utilization of soft and crowdsourced labels. For example, in datasets with soft and crowdsourced labels, adopting a lower binary threshold for sound events, such as 0.45 or 0.4, is more effective than using the standard threshold of 0.5 when training the model or creating strongly labeled datasets.
    

\end{enumerate}

The remainder of this paper is structured as follows: Section 2 describes the different types of label noise encountered in Sound Event Detection (SED). Section 3 then introduces the methods for generating noisy labels using both synthetic and real-life datasets. Section 4 details the experimental setup and describes the noise-robust functions evaluated. Finally, Section 5 presents the results of these experiments, along with theoretical analysis and discussion of their implications and significance.

\section{Label Noise Categories in SED}
\label{sec:noisy-label}

\begin{figure}[t]
    \centering
    \centerline{\includegraphics[width=8.5cm]{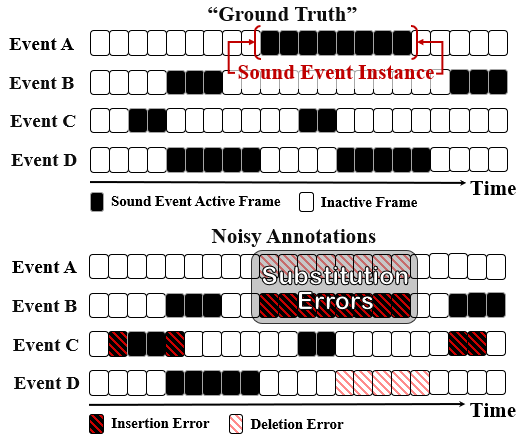}}
    \captionsetup{skip=1pt} 
    \caption{Ground truth and noisy annotated labels on the same audio clip. In the annotations, erroneous annotated frames are represented with red stripes. Incorrectly classified active is referred to 'insertion error,' and misclassified active is termed 'deletion error'. The continuous frames constitute a sound event instance.}
    \label{fig:ground_truth}
    \vspace{-2ex} 
\end{figure}

\begin{figure}[t]
    \centering
    \centerline{\includegraphics[width=8.5cm]{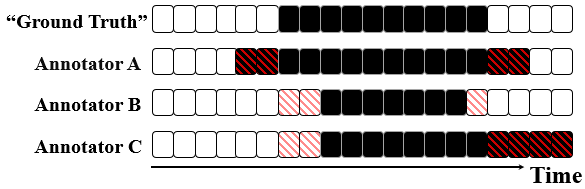}}
    \captionsetup{skip=1pt} 
    \caption{Subjective selection of boundaries by different annotators. Surplus and missing frames are marked with red stripes as insertion and deletion errors respectively.}
    \label{fig:subjective_label}
    \vspace{-2ex} 
\end{figure}

In SED, the audio clip is in general first divided into short time frames, such as 200 ms. Each frame is then analyzed using multi-label classification to determine the presence or absence of various sound event classes. An example of this segmentation is shown in Figure \ref{fig:ground_truth}, where frames containing sound events are marked as active, while others are marked as inactive. The continuous frames constitute a sound event instance, typically identified by the onset and offset times in the annotation files.
However, accurately labeling these frames is a challenge, primarily due to the intricate nature of acoustic signals and the subjective aspects of determining temporal boundaries. Drawing on the concept of error rates from \cite{c17}, we categorize label noise into three types, as illustrated at the frame level in the \textit{Noisy Annotations} part of Fig. \ref{fig:ground_truth}.


\begin{itemize}[parsep=0pt]

  \item \textbf{Deletion Label Noise:}
    This arises when active frames are erroneously labeled as inactive. For instance, the overlooked $Event\ D$ in Fig. \ref{fig:ground_truth}, which is active in the ground truth, illustrates deletion noise.

  \item \textbf{Insertion Label Noise:}
    This type of noise happens when frames are incorrectly marked as active for any class, although they are inactive in reality. An example is the mislabeled $Event\ C$ in Fig. \ref{fig:ground_truth}, which is absent in the ground truth.
    
  \item \textbf{Substitution Label Noise:}
    This occurs when a frame is wrongly annotated as a different class (e.g., $Event\ A$ labeled as $Event\ B$), as depicted in Fig. \ref{fig:ground_truth}. The substitution can be a combination of insertion and deletion: deletion of $Event\ A$ and insertion of $Event\ B$.

\end{itemize}


Given the subjective nature of selecting temporal boundaries for sound events, discrepancies often arise among annotators regarding their precise locations \cite{c18}. This type of label noise, stemming from individual interpretation differences, is addressed distinctly in our study. We refer to it as \textbf{Subjective Label Noise}, a concept further illustrated in Fig. \ref{fig:subjective_label}. Subjective label noise may include insertion, deletion, or both types of label noise.

\section{Constructing Datasets with Noisy Labels}
\label{sec:dataset}


We used the synthetic dataset Urban-SED \cite{c20} and the MAESTRO-Real \cite{c27}, a real-life dataset annotated with soft labels, to construct datasets with noisy labels.

\subsection{Synthetic Dataset: Urban-SED}

URBAN-SED comprises 10 sound event classes, with each audio sample having a duration of 10 seconds. As outlined in Section 2, we explored four label noise types for our experiments: deletion, insertion, substitution, and subjective noisy labels. To create these noisy labels, we modified the original URBAN-SED label files at the sound event instance level, as follows:

\begin{itemize}[parsep=0pt]
  \item \textbf{Deletion Label Noise:}
  
    This noise type was created by randomly removing sound event instances from the URBAN-SED's original label files, on a class-by-class basis. The degree of deletion was controlled by a parameter $deletion\_rate$, which varied from 0.0 to 0.5 in increments of 0.05, to control how many instances were removed. For example, there are $2683$ instances of $dog\_bark$ in the training dataset, and $1341$ event instances will be randomly removed with a deletion rate of 0.5. 

  \item \textbf{Insertion Label Noise:}
  
    For insertion noise, we randomly inserted additional sound event instances into the original URBAN-SED annotations, which were controlled by an $insertion\_rate$ (ranging from 0 to 1.0, with increments of 0.1), to determine the proportion of event instances added to each class. The inserted events' onset times were randomly set between 0 and 10 seconds, and their durations were aligned with the mean and standard deviation values of the respective event class in the original dataset.

  \item \textbf{Substitution Label Noise:}
  
    We considered the URBAN-SED's original labels as the ground truth and generated substitution noise by randomly changing the sound event class labels to one of the other nine classes. This was regulated by a $substitution\_rate$, varying from 0.0 to 0.5 (in 0.05 increments), to control the extent of substitution in each class. 
    
  \item \textbf{Subjective Noisy Label:}

    This type of noise was generated by perturbing the onset and offset times, randomly centered around the true value, by a variance factor called $overlap\_rate$. This factor quantifies the extent of overlap between the modified and original event durations, calculated as the overlap duration divided by the original event duration. It ranges from 0.5 to 1.0 with increments of 0.1.
    
\end{itemize} 

\subsection{Real-life Dataset: MAESTRO-Real}
The MAESTRO-Real dataset released by \cite{c27} provides temporally strong labels for sound events in real environments, with a unique feature of soft labels. These soft labels express the probability of the annotators on a scale from 0 to 1 that the event is present at that time instant, rather than the usual binary label of event present or not present.


By applying different a binary thresholds, we can transform these soft labels into various hard-label datasets. For instance, a hard-label dataset created with a 0.5 threshold can serve as our "ground truth". Datasets with thresholds below 0.5 will exhibit insertion noise, while those above 0.5 will show deletion noise. By dynamically adjusting the threshold within a specified range, we can simulate subjective noise, mimicking the variability in annotator judgment regarding event onset and offset times. In our study, we do not consider substitution noise due to the difficulty of discerning rules in real-life environments. 

We assume that hard labels derived at a 0.5 threshold represent the ground truth. The construction of insertion, deletion, and subjective label noise is as follows:

\begin{itemize}[parsep=0pt]
    \item \textbf{Insertion and Deletion Label Noise}: 
    
        Generating multiple hard-label datasets with thresholds ranging from 0.1 to 0.9 in 0.05 steps. Thresholds below 0.5 incorporate increasing levels of insertion noise, while those above 0.5 include increasing degrees of deletion noise.
    
    \item \textbf{Subjective Label Noise}: 
    
        To create hard-label datasets with subjective noise, we introduce a relaxation factor $\Omega$, defining a binary threshold range of [0.5-$\Omega$, 0.5+$\Omega$]. For example, with a relaxation factor of 0.1, the range would be [0.4, 0.6]. Each data sample is then binarized using a randomly selected threshold within this range, resulting in a dataset reflecting subjective labeling noise. The range of $\Omega$ varies from $0$ to $0.45$, with a step of $0.05$.
\end{itemize} 
\section{Experimental setup \& Noise-Robust Loss Functions}
\label{sec:setups}
\begin{table*}[t!]
    \centering
    \captionsetup{skip=1pt} 
    \caption{The $w_{active}$ and $w_{inactive}$ of each loss function in SRL}
    \begin{tabular}{l | c | c | c |c }
        \hline
        \textbf{ } & \textbf{$\mathcal{L}_{bce}$ }  & \textbf{$\mathcal{L}_{boot}$}  & \textbf{$\mathcal{L}_{smooth}$}  & \textbf{$\mathcal{L}_{gce}$} \\ 
        \hline
        $w_{active}$ & $-\mathbf{y} \cdot \log(\mathbf{p})$ &  $-\Tilde{\mathbf{y}}\cdot \log(\mathbf{p})$ & $-\mathbf{y_{smooth}}\cdot \log(\mathbf{p})$  & $\mathbf{y} \cdot \frac{1 - \mathbf{p}^q}{q}$  \\
        \hline
        $w_{inactive}$ & $-(\mathbf{1} - \mathbf{y}) \cdot \log(\mathbf{1} - \mathbf{p})$ &  $-(\mathbf{1} - \Tilde{\mathbf{y}}) \cdot \log(\mathbf{1} - \mathbf{p})$  & $-(\mathbf{1} - \mathbf{y_{smooth}}) \cdot \log(\mathbf{1} - \mathbf{p})$ &  $(\mathbf{1} - \mathbf{y}) \cdot \frac{1 - (1 - \mathbf{p})^q}{q}$\\        
        \hline
    \end{tabular}
    \label{table: srl_illus}
    \vspace{-2ex} 
\end{table*}
We use the popular CRNN architecture for SED tasks \cite{c8} to investigate the impact of label noise on system performance with both synthetic and real-life datasets. Each experimental setup runs at least 5 times and takes the average value of evaluation metrics as the final results. The training and validation datasets contain noisy labels while the testing dataset adopts the "ground-truth" labels to evaluate the impact of label noise on SED system performance. 


\subsection{SED Model \& Evaluation Metrics}
We explore the baseline system from the DCASE 2023 Challenge Task 4B by adding a sigmoid layer as the final output and changing the epoch number to 70 to balance the model accuracy and training time. The baseline system is a standard CRNN with three convolutional blocks and one bidirectional GRU layer of size 32. Each convolutional block consists of a single 2D convolution with 128 filters and $3\times3$ kernels followed by batch normalization, rectified linear unit, max pooling, and dropout layers. The model utilized mel-band energies as input, extracted with a hop length of 200 ms and 64 mel filter banks. For more detailed parameters settings can refer to the GitHub baseline systems\textsuperscript{1}.

\footnotetext[1]{\href{https://github.com/marmoi/dcase2023\_task4b\_baseline}{https://github.com/marmoi/dcase2023\_task4b\_baseline}}

The outputs are evaluated by comparing them with the "ground-truth" labels. For synthetic Urban-SED, the "ground-truth" labels are the original labels of the test datasets in Urban-SED. For real-life MAESTRO-Real, the hard labels derived at a 0.5 threshold represent the ground truth as we mentioned in Section 3.2. The sigmoid outputs are probabilities between 0 and 1, so a threshold of 0.5 is applied to each output to determine the predicted event class activity independently.  System evaluation is based on the 1s segment-based F1-score and error rate \cite{c17}, including micro-average F1 score (F1\_m), micro-average error rate (ER), and macro-average F1 score (F1\_M). Since the F1 score offers a more intricate and balanced measure of a model's performance, we prefer to enhance the F1 score while keeping the ER at an acceptable level to ensure overall reliability.

\subsection{Noise-robust Loss Functions for Classification}
Training a deep network is based on updating the network weights to minimize a loss function that measures the divergence between the network predictions and the target labels. If the target labels contain errors, the update of network weights can be suboptimal thus hindering the model convergence. In these cases, the loss functions need to be robust against noisy labels.

We choose three popular noise-robust loss functions that have been used in multi-class classification tasks and apply them to multi-label classification problems such as SED. For the sake of simplicity, we omit the procedure of summing the loss across all training data frames and subsequently computing the average within a training batch. Instead, we define the loss computation at the individual frame level. The commonly-used loss function for multi-label classification is the binary cross entropy loss (BCE), as shown in the formula (\ref{eq: bce_loss}):
\vspace{-1ex} 
\begin{equation}
\label{eq: bce_loss}
\mathcal{L}_{bce} = - \mathbf{y} \cdot \log(\mathbf{p}) - (\mathbf{1} - \mathbf{y}) \cdot \log(\mathbf{1} - \mathbf{p})
 \end{equation}%
where $\mathbf{y} \in \{0,1\}^{M}$ represent the target multi-hot labels of $M$ sound event classes in a frame, while $\mathbf{p} \in [0,1]^{M}$ represent the network predictions (the predicted class probabilities) for the corresponding frame.

The first noise-robust loss function we choose is referred to as soft bootrapping \cite{c28} (BootStrap), called $\mathcal{L}_{boot}$. The idea of $\mathcal{L}_{boot}$ is to update the target labels based on the current state of the model. In particular, the updated target labels are a convex combination of the current model's outputs and the (potentially noisy) target label, as expressed by formula (\ref{eq: soft_bootstrap}) and formula (\ref{eq: soft_target}):
\begin{equation}
\label{eq: soft_bootstrap}
\mathcal{L}_{boot} = - \mathbf{\Tilde{y}}\cdot \log(\mathbf{p}) - (\mathbf{1} - \mathbf{\Tilde{y}}) \cdot \log(\mathbf{1} - \mathbf{p})
 \end{equation}
 

 \begin{equation}
\label{eq: soft_target}
\mathbf{\Tilde{y}} = \beta \cdot \mathbf{y} + (1-\beta) \cdot \mathbf{p}
 \end{equation}
where $\beta \in (0,1]$ is the blending factor, balancing the contribution of the original target label and the model's prediction. In the experiments, $\beta$ varies from 0.1 to 1.0 in increments of 0.1 to select the relatively optimal value.

The second technique we adopt is label smoothing (LabelSmooth), which has been proven to not only improve the generalization of the model but also mitigate the impact of label noise \cite{c23}, \cite{c24}. This method softens hard labels (0 or 1) to reduce the model's sensitivity to noise. We extend the original LabelSmooth loss to multi-label classification. The formula is similar to BCE except that the target label is modified by a smoothing parameter, as depicted in formulas (\ref{eq: smooth_loss}) and (\ref{eq: smooth_y}):
\begin{equation}
\label{eq: smooth_loss}
\mathcal{L}_{smooth} = - \mathbf{y_{smooth}}\cdot \log(\mathbf{p}) - (\mathbf{1} - \mathbf{y_{smooth}}) \cdot \log(\mathbf{1} - \mathbf{p})
 \end{equation}


\begin{equation}
\label{eq: smooth_y}
\mathbf{y_{smooth}}\ = \mathbf{y} \cdot (1 - \alpha) + \alpha / 2.
 \end{equation}
where $\alpha \in [0,1)$ is the smoothing parameter. To find a more optimized $\alpha$, the value of $\alpha$ starts at 0.05 and doubles with each step, resulting in: [0.05, 0.1, 0.2, 0.4, 0.8].


The third noise-robust loss function is a generalization of categorical cross-entropy loss (CCE) and the mean absolute error (MAE) proposed in \cite{c29}, called generalized cross-entropy loss (GCE). As stated by the authors, GCE loss combines the advantages of both CCE and MAE and can be readily applied with any existing DNN architecture and algorithm, while yielding good performance in a wide range of noisy label scenarios. We have extended the original GCE loss, initially designed for multi-class classification, for binary classification, as shown in formula (\ref{eq: gce_loss})
\begin{equation}
\label{eq: gce_loss}
\mathcal{L}_{gce}  = \mathbf{y} \cdot \frac{1 - \mathbf{p}^q}{q} + (\mathbf{1} - \mathbf{y}) \cdot \frac{1 - (\mathbf{1} - \mathbf{p})^q}{q} 
\end{equation}
where $q \in (0, 1]$, determines the behavior of the loss function. As $q$ approaches 1, the loss function behaves more like the MAE, and as $q$ approaches 0, it behaves more like the BCE. The value of $q$ varies from 0.1 to 0.9 in increments of 0.1 to select the optimal value within this range.

\subsection{Simple Reweighting Loss Function}
Since deletion label errors are more detrimental than insertion label errors based on the findings of Section 5, we realize that the loss functions that are designed to alleviate the data imbalance between foreground sound events and background sounds should be effective in mitigating the impact of deletion noisy labels in SED. The reason is that, during model training, reducing the weight of inactive frames is essentially equivalent to diminishing the impact of deletion label errors, which aligns with the core principle of dealing with noisy labels. To verify this inference, we extend the \textit{Simple Reweighting Loss (SRL)} method initially proposed by \cite{c31} to all the loss functions described above. This method reweights the active and inactive frames separately during model training, as shown in formula (\ref{eq: srl}):

\begin{equation}
\label{eq: srl}
\mathcal{L}_{srl} = \gamma \cdot w_{active} + \xi \cdot w_{inactive}
 \end{equation}
where $w_{active}$ and $w_{inactive}$ represent the two components of each loss function as identified in Table \ref{table: srl_illus}. $\gamma \in  [0, \infty)$ and $\xi \in  [0, \infty)$ are the reweighting factors. In this work, we set $\gamma$ as $1.0$ and vary $\xi$. For $\xi=1.0$ we obtain the original (unweighted) loss functions.


\section{Results and Analysis}
\label{sec: results}

The Binary Cross-Entropy (BCE) loss serves as a baseline for comparison. Initially, experiments are conducted to determine the "optimal" hyperparameters for three popular noise-robust loss functions. Subsequently, these optimal hyperparameters are utilized to compare the performance of each noise-robust loss function with the baseline. Finally, we implement the Simple Reweighting Loss (SRL) paradigm across all baseline and noise-robust loss functions. This step is undertaken to validate our hypothesis that reducing the weight of inactive sound event frames (background sound) can effectively mitigate the impact of noisy labels and enhance model robustness. These procedures are applied to both synthetic and real-life datasets. Detailed results and analyses are presented separately for the synthetic and real-life datasets.

\subsection{Synthetic Dataset}

\begin{figure*}[t!]

    \begin{minipage}[b]{1.0\linewidth}
        \centering
        \centerline{\includegraphics[width=\textwidth]{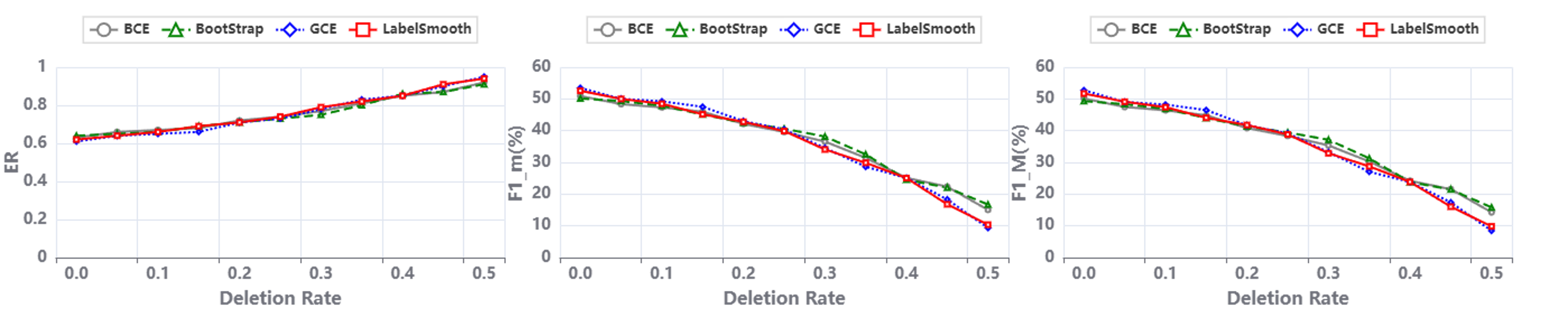}}
    \end{minipage}
    \captionsetup{skip=1pt} 
    \caption{Impact of Deletion Label Noise on SED Performance in Synthetic Datasets, Evaluated by ER, F1\_m, and F1\_M}
    \label{fig: syn_deletion}
%
\end{figure*}

\begin{figure*}[t!]

    \begin{minipage}[b]{1.0\linewidth}
        \centering
        \centerline{\includegraphics[width=\textwidth]{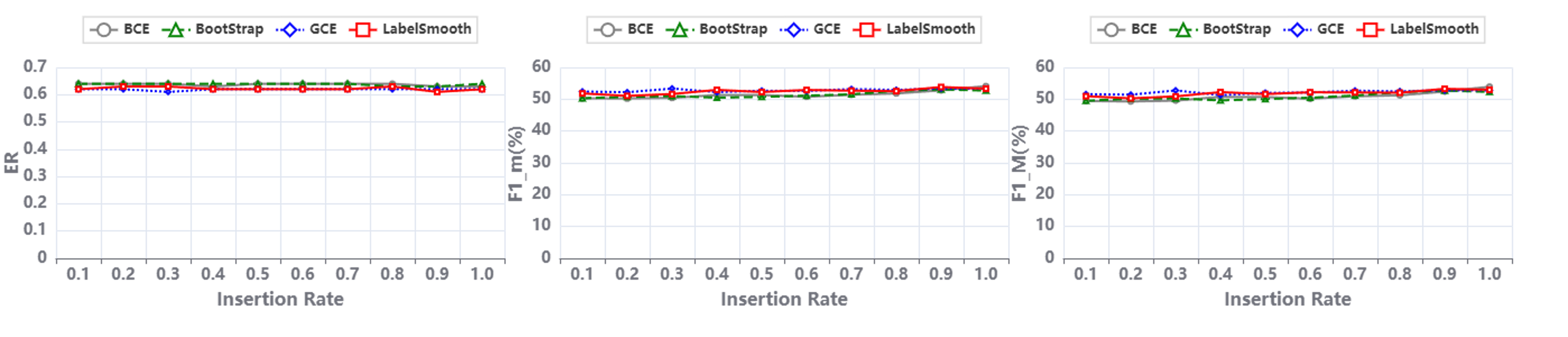}}
    \end{minipage}
    \captionsetup{skip=1pt} 
    \caption{Impact of Insertion Label Noise on SED Performance in Synthetic Datasets, Evaluated by ER, F1\_m, and F1\_M}
    \label{fig: syn_insertion}
%
\end{figure*}

\begin{figure*}[t!]

    \begin{minipage}[b]{1.0\linewidth}
        \centering
        \centerline{\includegraphics[width=\textwidth]{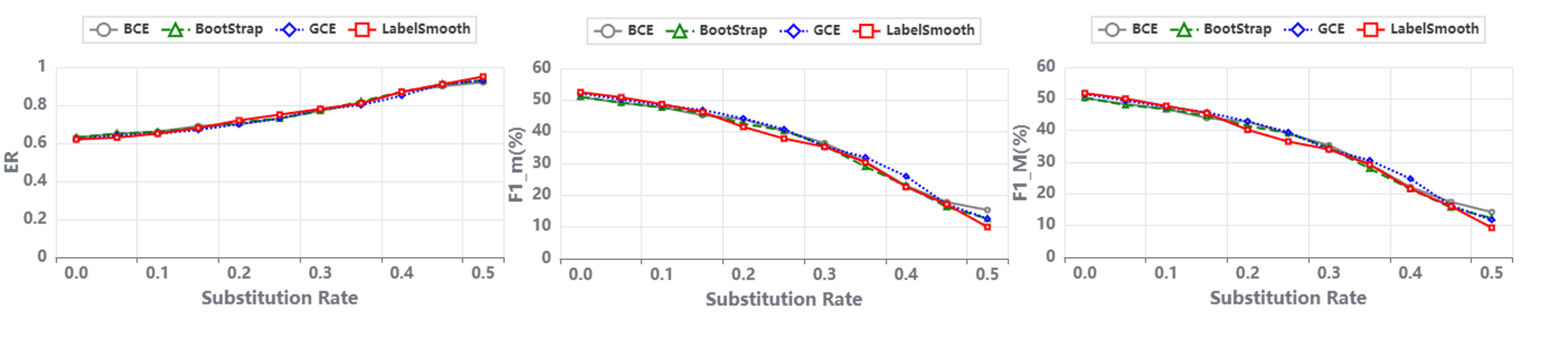}}
    \end{minipage}
    \captionsetup{skip=1pt} 
    \caption{Impact of Substitution Label Noise on SED Performance in Synthetic Datasets, Evaluated by ER, F1\_m, and F1\_M}
    \label{fig: syn_substitute}
%
\end{figure*}

\begin{figure*}[t!]

    \begin{minipage}[b]{1.0\linewidth}
        \centering
        \centerline{\includegraphics[width=\textwidth]{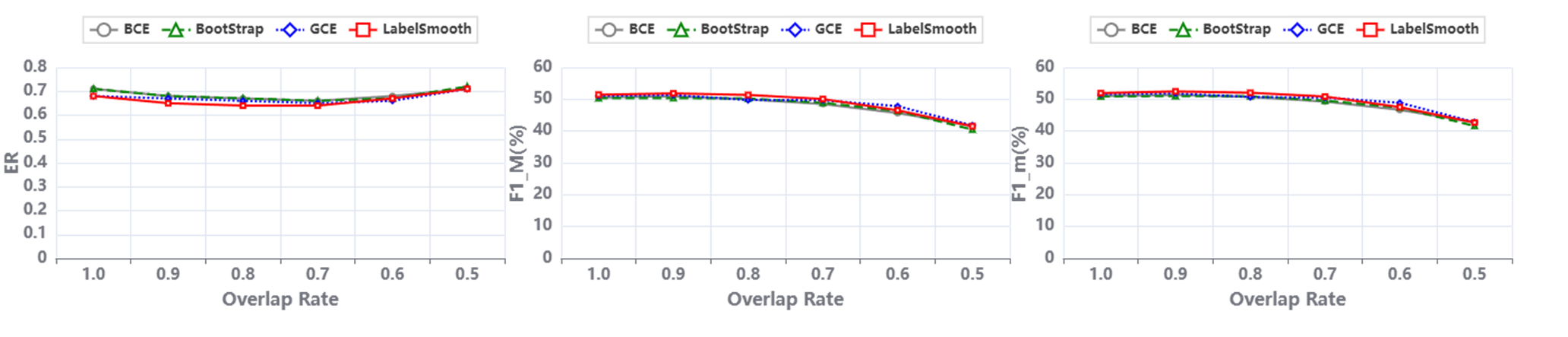}}
    \end{minipage}
    \captionsetup{skip=1pt} 
    \caption{Impact of Subjective Label Noise on SED Performance in Synthetic Datasets, Evaluated by ER, F1\_m, and F1\_M}
    \label{fig: syn_subjective}
%
\end{figure*}


\begin{table*}[ht]
\centering
\caption{Average SED performance metrics across deletion, insertion, substitution, and subjective label noises for various $\xi$ values in \textbf{synthetic dataset}.
}
\label{table: syn_srl}
\begin{tabular}{|c|ccc|ccc|ccc|ccc|}
\hline
\multirow{2}{*}{$\xi$} & \multicolumn{3}{c|}{BCE} & \multicolumn{3}{c|}{Bootstrap} & \multicolumn{3}{c|}{LabelSmooth} & \multicolumn{3}{c|}{GCE} \\
& ER & F1\_m & F1\_M & ER & F1\_m & F1\_M & ER & F1\_m & F1\_M & ER & F1\_m & F1\_M\\
\hline
\hline
1.0	&	\textbf{0.714}	&	42.41	&	41.547	&	\textbf{0.714}	&	42.337	&	41.481	&	0.709	&	42.39	&	41.514	&	\textbf{0.705}	&	42.893	&	41.964	\\
\rowcolor{yellow!50}
0.5	&	0.735	&	\textbf{50.663}	&	50.469	&	0.735	&	\textbf{50.612}	&	\textbf{50.417}	&	\textbf{0.708}	&	\textbf{51.03}	&	50.79	&	0.711	&	\textbf{51.437	}&	\textbf{51.161}	\\
0.25	&	1.179	&	50.024	&	\textbf{50.521}	&	1.188	&	49.937	&	50.414	&	1.126	&	50.651	&	\textbf{51.103}	&	1.15	&	50.595	&	51.07	\\
0.125	&	2.196	&	43.301	&	43.826	&	2.184	&	43.435	&	43.958	&	2.196	&	43.476	&	43.978	&	2.222	&	43.355	&	43.918	\\
0.0625	&	3.633	&	35.794	&	36.082	&	3.593	&	35.972	&	36.282	&	3.699	&	35.43	&	35.676	&	3.671	&	35.679	&	35.974	\\
\hline
\end{tabular}
\end{table*}

\begin{table*}[ht]
\centering
\caption{Average SED performance metrics across deletion, insertion, and subjective label noises for various $\xi$ values in the \textbf{real-life dataset}.
}
\label{table: real_srl}
\begin{tabular}{|c|ccc|ccc|ccc|ccc|}
\hline
\multirow{2}{*}{$\xi$} & \multicolumn{3}{c|}{BCE} & \multicolumn{3}{c|}{Bootstrap} & \multicolumn{3}{c|}{LabelSmooth} & \multicolumn{3}{c|}{GCE} \\
& ER & F1\_m & F1\_M & ER & F1\_m & F1\_M & ER & F1\_m & F1\_M & ER & F1\_m & F1\_M\\
\hline
\hline
1.0	&	\textbf{0.759}	&	52.03	&	20.608	&	\textbf{0.757}	&	51.615	&	20.126	&	\textbf{0.761}	&	53.197	&	22.601	&	\textbf{0.761}	&	51.745	&	20.214	\\
\rowcolor{yellow!50}
0.5	&	0.78	&	57.32	&	25.972	&	0.78	&	57.006	&	25.313	&	0.804	&	\textbf{57.499}	&	28.078	&	0.777	&	57.384	&	25.535	\\
\rowcolor{green!50}
0.25	&	0.918	&	\textbf{58.258}	&	30.149	&	0.923	&	\textbf{57.883}	&	29.994	&	0.981	&	56.996	&	30.54	&	0.921	&	\textbf{57.968}	&	30.103	\\
0.125	&	1.213	&	55.545	&	\textbf{31.268}	&	1.213	&	55.484	&	\textbf{31.184}	&	1.291	&	53.892	&	\textbf{30.689}	&	1.215	&	55.371	&	\textbf{31.23}	\\
0.0625	&	1.558	&	52.395	&	30.87	&	1.557	&	52.478	&	30.877	&	1.698	&	50.326	&	29.574	&	1.562	&	52.226	&	30.795	\\
\hline
\end{tabular}
\end{table*}


\begin{figure*}[t!]

    \begin{minipage}[b]{1.0\linewidth}
        \centering
        \centerline{\includegraphics[width=\textwidth]{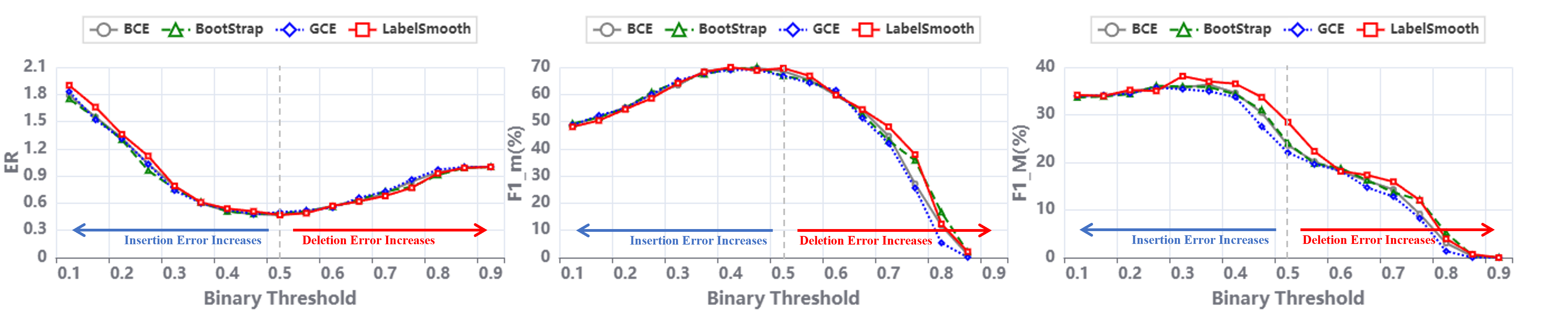}}
    \end{minipage}
    \captionsetup{skip=1pt} 
    \caption{Impact of Insertion and Deletion Label Noise on SED Performance in Real-life Datasets, Evaluated by ER, F1\_m, and F1\_M}
    \label{fig: real_insert_del}
    \vspace{-2ex} 
\end{figure*}

\begin{figure*}[t!]

    \begin{minipage}[b]{1.0\linewidth}
        \centering
        \centerline{\includegraphics[width=\textwidth]{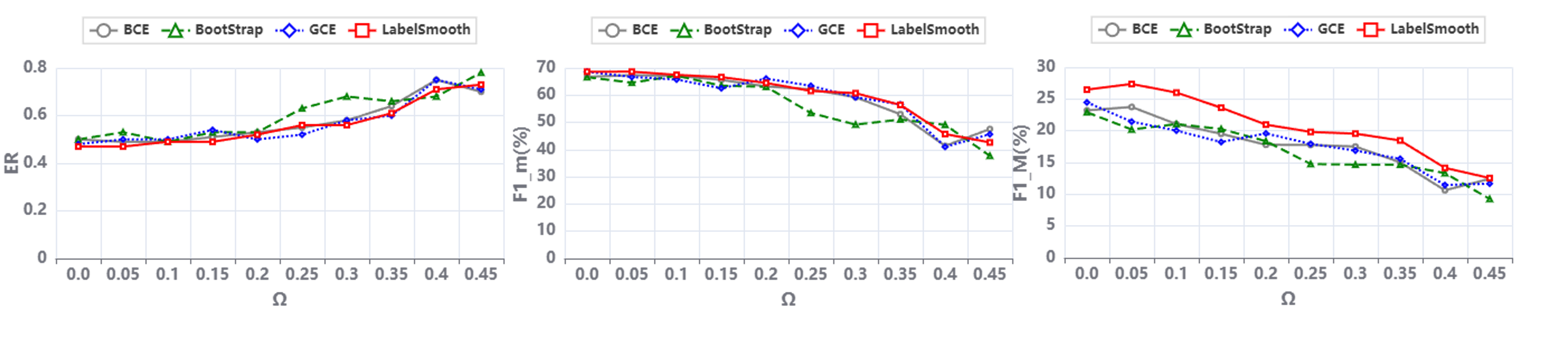}}
    \end{minipage}
    \captionsetup{skip=1pt} 
    \caption{Impact of Subjective Label Noise on SED Performance in Real-life Datasets, Evaluated by ER, F1\_m, and F1\_M}
    \label{fig: real_subject}
%
\end{figure*}

\begin{figure}[t]
    \centering
    \centerline{\includegraphics[width=8.5cm]{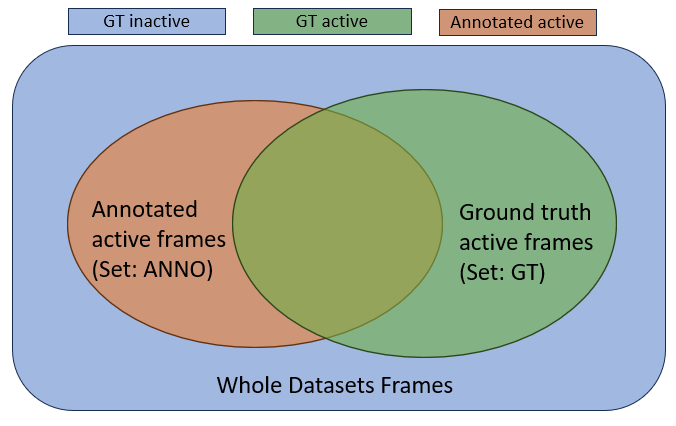}}
    \captionsetup{skip=1pt} 
    \caption{Illustration of Set Perspective on Annotation for One Sound Event in the Dataset }
    \label{fig: set_annotation}
    \vspace{-2ex} 
\end{figure}

\begin{figure}[t]
    \centering
    \centerline{\includegraphics[width=8.5cm]{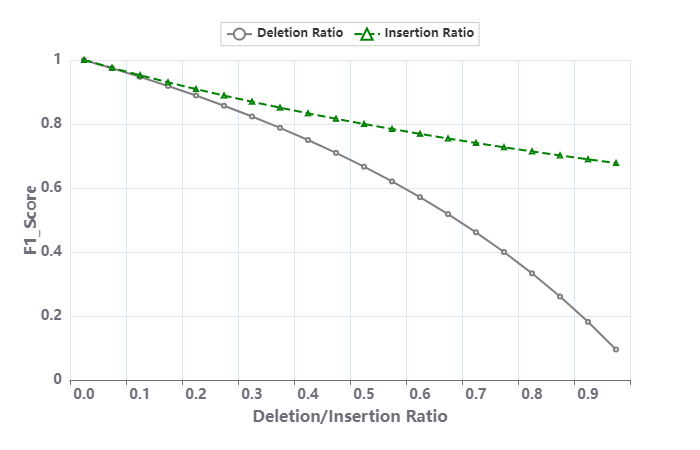}}
    \captionsetup{skip=1pt} 
    \caption{Impact of Deletion/Insertion Ratio on F1-Score }
    \label{fig: theo_res}
    \vspace{-2ex} 
\end{figure}

\begin{figure}[t]
    \centering
    \centerline{\includegraphics[width=8.5cm]{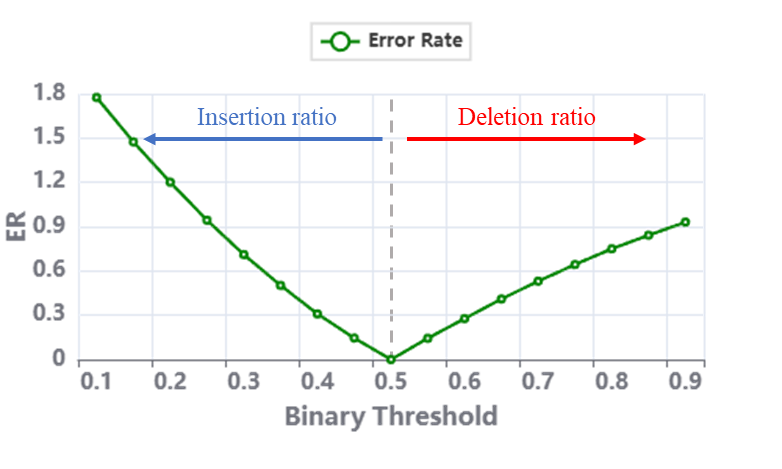}}
    \captionsetup{skip=1pt} 
    \caption{Theoretical Error Rate of Real-life Datasets with Different Binary Threshold}
    \label{fig: theo_er}
    \vspace{-2ex} 
\end{figure}

\begin{figure*}[t!]

    \begin{minipage}[b]{1.0\linewidth}
        \centering
        \centerline{\includegraphics[width=\textwidth]{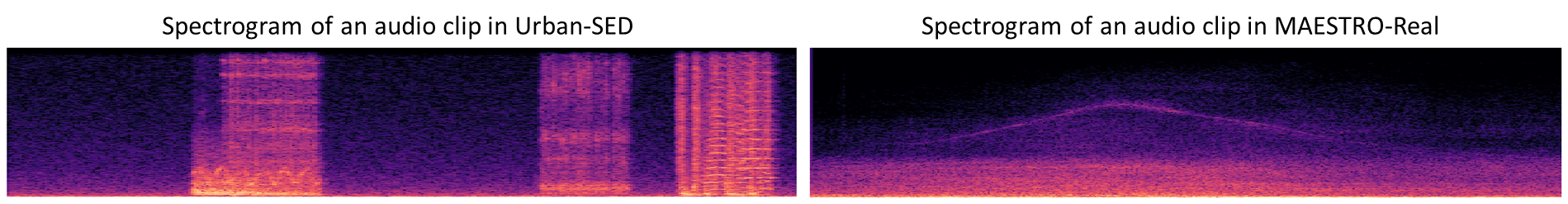}}
    \end{minipage}
    \captionsetup{skip=1pt} 
    \caption{Expample of Spectrogram Visualization of Synthetic and Real-life Soundscapes}
    \label{fig: spectrogram}
%
\end{figure*}



Through hyperparameter grid search, we determined the optimal values: $\beta=0.9$, $\alpha=0.4$, and $q=0.5$ as the default for $\mathcal{L}_{boot}$, $\mathcal{L}_{smooth}$, and $\mathcal{L}_{gce}$, respectively. With these optimal hyperparameters identified, we now examine the performance of each loss function from Section 4.2 under various noisy label conditions.

\subsubsection{Deletion Label Noise}

Fig. \ref{fig: syn_deletion} illustrates the trend of performance degradation as the $deletion\_rate$ increases. It is evident that the model's performance deteriorates rapidly with higher deletion rates, indicating a significant impact of deletion label noise on SED model performance. Furthermore, all the loss functions exhibit a similar trend, suggesting that noise-robust loss functions effective in classification tasks may not be suitable for addressing deletion noise in SED.

\subsubsection{Insertion Label Noise}

Fig. \ref{fig: syn_insertion} depicts the impact of insertion label noise on the performance of SED models. As the $insertion\_rate$ increases, there is no significant degradation in performance. This stability suggests that insertion label noise may have a minimal impact on SED performance. Further explanation is provided in Section 5.4.

\subsubsection{Substitution Label Noise}
The impact of substitution label noise is very similar to that of deletion label noise, as shown in Fig. \ref{fig: syn_substitute}. Since substitution label noise can be considered a combination of insertion and deletion noises, such results are unsurprising based on previous findings that insertion errors have a minimal impact on SED performance. Moreover, from the perspective of active frames, substitution label noise corresponds to mislabeling errors in classification tasks. This reinforces the conclusion that noise-robust loss functions used in classification to handle corrupted labels do not offer significant advantages in SED scenarios.

\subsubsection{Subjective Label Noise}

Fig. \ref{fig: syn_subjective} illustrates the performance trend as the $overlap\_rate$ decreases. It is observable that at lower $overlap\_rate$, the performance of the SED model is adversely affected, though to a lesser extent than with deletion label noise, and more substantially than with insertion label noise. The primary cause of this phenomenon is that subjective label noise is introduced by adjusting the start and end times of independent sound events. Although this creates some deletion errors, they are less severe than entirely removing the sound event.

\subsubsection{Investigating the Impact of SRL}

The experimental results, averaged across all four noise types and shown in Table \ref{table: syn_srl}, demonstrate that halving the weights of inactive frames markedly improves model performance for all loss functions. For example, with the \textit{GCE} loss, reducing $\xi$ from 1.0 to 0.5 improves the F1\_m and F1\_M scores from 42.893\% and 41.964\% to 51.437\% and 51.161\% respectively, accompanied by a slight increase in Error Rate (ER) from 0.705 to 0.711. This result confirms our inference that decreasing the weight of inactive frames can mitigate the impact of deletion label errors, as insertion label errors have a negligible impact on model performance.

\subsection{Real-life Dataset}


Similar to the synthetic dataset analysis, we first determine the optimal hyperparameters for three well-known noise-robust loss functions in the real-life dataset. The optimal hyperparameter settings differ slightly from those in the synthetic dataset. For the real-life dataset, we establish $\beta=0.9$, $\alpha=0.6$, and $q=0.1$ as the default values for $\mathcal{L}_{boot}$, $\mathcal{L}_{smooth}$, and $\mathcal{L}_{gce}$, respectively.

\subsubsection{Insertion and Deletion Label Noise}

Fig. \ref{fig: real_insert_del} shows the model performance trend as the binary threshold changes. In terms of ER, the impact of insertion label noise is more pronounced than that of deletion label noise, which will be explained in the theoretical analysis. However, for the F1 scores, particularly F1\_M, the impact of deletion noise far outweighs that of insertion noise, aligning with the findings from the synthetic dataset. The three noise-robust loss functions did not significantly outperform the baseline \textit{BCE} loss, except for \textit{LabelSmooth} loss, which exhibited a noticeable improvement in F1 scores. The \textit{LabelSmooth} loss reduces the impact of noisy labels to some extent by mapping active and inactive labels from 1 and 0 to $1-\alpha/2$ and $\alpha/2$. Due to the unique intra-class imbalance in SED \cite{c34}, where the number of inactive frames is often much greater than the number of active frames, compressing the label values of all active and inactive frames more effectively suppresses the influence of inactive frame labels. This, in turn, more effectively suppresses the influence of deletion noisy labels.

\subsubsection{Subjective Label Noise}

Fig. \ref{fig: real_subject} illustrates that as $\Omega$ (subjective label noise rate) increases, the model's performance suffers, marked by a rise in ER and a decline in F1 scores. This trend highlights the substantial impact of subjective label noise on model performance. However, its effect is less severe compared to deletion label noise, as evidenced by the F1 scores not dropping to zero but maintaining a relatively high level. This can be attributed to the fact that while some annotators may overlook parts of a sound event, others recognize them, thereby mitigating the impact of deletion errors to some extent. Again, the \textit{LabelSmooth} loss function demonstrates superior performance in the F1\_M metric compared to the other loss functions.

\subsubsection{Investigating the Impact of SRL}

The impact of SRL on real-life datasets is consistent with findings from synthetic datasets, as shown in Table \ref{table: real_srl}. Taking the \textit{BCE} loss as an example, halving the weight of non-event background sounds leads to an increase in average F1\_m from 52.03\% to 57.32\% and F1\_M from 20.608\% to 25.972\%, with only a minor increase in ER from 0.759 to 0.78. To further improve the F1\_M score, reducing the weight of non-event background sounds even more, for instance, changing $\xi$ from 1.0 to 0.25, could raise the F1\_M to 30.149\%, though this would result in an ER increase to 0.918.

\subsection{Theoretical Analysis}
To analyze the impact of deletion and insertion label errors theoretically, suppose we have an ideal model that predicts results perfectly matching the annotations. Now, let's examine the impact of insertion and deletion noisy labels on SED performance. Figure \ref{fig: set_annotation} illustrates the annotations from a set perspective, where $GT$ represents the ground truth active frames and $ANNO$ represents the annotated active frames. Let $T_{act}$, $T_{inact}$, $T_{del}$, and $T_{insert}$ denote the total number of ground truth active frames, ground truth inactive frames, deleted active frames, and inserted active frames, respectively, where $T_{act} = |GT|$, $T_{inact} = |GT^c|$, $T_{del} = |GT \setminus ANNO|$, and $T_{insert} = |ANNO \setminus GT|$. The deletion and insertion ratios can then be defined as formula (\ref{eq: del_ratio}) and (\ref{eq: insert_ratio}): 

\begin{equation}
\label{eq: del_ratio}
R_{del} = T_{del} / T_{act}
\end{equation}

\begin{equation}
\label{eq: insert_ratio}
R_{insert} = T_{insert} / T_{act}
\end{equation}

Let's analyze the impact of deletion error labels by first setting the insertion ratio to 0. Precision (P), recall (R), F1-score, and error rate are calculated as follows:

\begin{equation}
\label{eq: recall_del}
R = 1 - R_{del}
\end{equation}

\begin{equation}
\label{eq: precise_del}
P = 1.0
\end{equation}

\begin{equation}
\label{eq: f1_del}
F1 = \frac{2PR}{P + R} = \frac{2\cdot (1 - R_{del})}{2 - R_{del}} 
\end{equation}

\begin{equation}
\label{eq: er_del}
ER = R_{del}
\end{equation}

Similarly, the impact of insertion error labels can be analyzed by setting the deletion ratio to 0, as follows:

\begin{equation}
\label{eq: recall_insert}
R = 1.0
\end{equation}

\begin{equation}
\label{eq: pre_insert}
P = \frac{1}{1 + R_{insert}}
\end{equation}

\begin{equation}
\label{eq: f1_insert}
F1 = \frac{2PR}{P + R} = \frac{2}{2 + R_{insert}} 
\end{equation}

\begin{equation}
\label{eq: er_insert}
ER = R_{insert}
\end{equation}

The impact of deletion ratio and insertion ratio on the F1-score can be demonstrated in Fig. \ref{fig: theo_res}. As we can see, deletion errors are more detrimental than insertion errors theoretically when comparing the F1-score. We should also note that since the insertion ratio can exceed 1, whereas the deletion ratio is limited to the range [0,1], severe insertion noise can lead to a more significant error rate (ER) due to insertion errors. Fig. \ref{fig: theo_er} demonstrates the theoretical variations of ER when changing the binary threshold in the real-life dataset, which is consistent with the ER variations shown in Fig. \ref{fig: real_insert_del}.

\subsection{Discussion and Limitations}




Both theoretical analysis and our experimental results conclude that deletion errors significantly impact system performance more than insertion errors in datasets with noisy labels. Additionally, the annotation experiments by \cite{c33} indicate that annotators tend to produce more deletion errors than insertion errors. These insights are crucial for improving data annotation and utilization. To minimize deletion errors, it may be beneficial to allow for a slight increase in insertion errors during annotation. For example, extending the start and end times of sound events can be effective when precise determination is challenging. Furthermore, in datasets with soft and crowdsourced labels, adopting a lower binary threshold for sound events, such as 0.45 or 0.4, is more effective than using the standard threshold of 0.5.


Moreover, unlike classification tasks that involve only classes, SED requires identifying sound event classes and their precise timings. Unlike simple clip-level classification, SED segments an audio clip into frames for multi-label analysis of each frame, followed by post-processing to finalize predictions. This process faces the unique challenge of intra-class imbalance. Events of interest often last significantly shorter than the background, leading to a disparity between active and inactive frames—a situation that minimally impacts standard classification but critically affects SED. Our experiments demonstrate that while noise-robust loss functions are effective for classification tasks, they are less effective in SED due to this imbalance.

Since intra-class imbalance plays a critical role in SED performance, deletion errors increase this imbalance by removing active frames, leading to a swift decline in model accuracy. In contrast, insertion noise retains more correct active frames than deletion noise, minimally impacting intra-class imbalance and causing no significant performance degradation as the insertion rate increases. This is why a loss function designed to address data imbalance can help improve resilience to label noise, particularly deletion noise. By diminishing the influence of inactive frames, such a loss function can directly mitigate the adverse effects of deletion errors during model training.

We also notice that insertion noise has a negligible effect on URBAN-SED but a non-negligible effect on MAESTRO-Real. We believe this difference mainly stems from the characteristics of the datasets. As shown in Fig. \ref{fig: spectrogram}, URBAN-SED, being randomly synthesized, exhibits distinct boundaries in the spectrogram, whereas the real dataset MAESTRO-Real has a smoother spectrogram. These distinct boundary features in URBAN-SED make the model almost unaffected by insertion noise.

However, this study has limitations due to our hardware constraints. We only tested the classic CRNN model and three popular noise-robust loss functions. Advanced models, like transformer-based architectures, and other noise-robust approaches were not explored. Our findings should encourage further research in these areas, contributing to the progress of the SED field.

\section{Conclusion}

In summary, this study provides the first in-depth analysis of different noisy label types on SED. We discovered that deletion label noise significantly affects SED performance more than insertion noise. Our experiments reveal that loss functions for standard classification are less effective for SED, whereas those addressing data imbalance between foreground and background sounds show promise. For instance, halving background sound weights in datasets notably improved macro-F1 and micro-F1 scores. These insights are crucial for developing noise-robust SED systems and highlight the importance of tailored approaches for handling noisy labels, especially in the use of soft and crowdsourced labels for dataset development.




\bibliographystyle{named}
\bibliography{ijcai24}

\end{document}